# Correlated multi-electron dynamics in ultrafast laser pulse - atom interactions

A. Rudenko[1], K. Zrost[1], B. Feuerstein[1], V.L.B. de Jesus[1,2], C.D. Schröter[1], R. Moshammer[1], and J. Ullrich[1]

[1] *Max-Planck-Institut für Kernphysik, D-69029 Heidelberg, Germany*
[2] *Centro Federal de Educação Tecnológica de Química de Nilópolis/RJ, Rua Lucio Tavares 1045, Centro - Nilópolis - 26530-060, Rio de Janeiro, Brazil*

We present the results of the detailed experimental study of multiple ionization of Ne and Ar by 25 and 7 fs laser pulses. For Ne the highly correlated "instantaneous" emission of up to four electrons is triggered by a recollisional electron impact, whereas in multiple ionization of Ar different mechanisms, involving field ionization steps and recollision-induced excitations, play a major role. Using few-cycle pulses we are able to suppress those processes that occur on time scales longer than one laser cycle.

Since the first experimental observation of doubly charged ions produced by intense laser fields [1], many-electron dynamics in laser-induced ionization has become one of the hottest topics in the field of laser-matter interactions. It was soon found out [2] that for linearly polarized light the yields of doubly and multiply charged ions exceed those predicted for independent successive removals of two or more electrons by many orders of magnitude. The origin of this enhancement (which was called 'non-sequential' multiple ionization) remained controversial until 2000 when the first measurements of recoil ion momentum distributions [3,4] along with the suppression of the enhancement for elliptically polarized light (see, e.g [5]) provided convincing evidences in favour of the so-called 'recollision' model [6]. As illustrated in Fig. 1, in this model an electron that tunnels out of the atom is then driven back to its parent ion and knocks out another electron. This mechanism was found to be the only one compatible with the characteristic double peak structure and the minimum at zero in the ion-momentum distributions along the field polarization ('longitudinal') direction observed in the experiments. Indeed, if an electron or ion is born at the maximum of the electric field, it acquires essentially zero drift momentum and the final momentum along the field polarization direction is small. Therefore, in the case of single ionization or sequential tunneling of several electrons the ion momentum distribution displays a maximum at zero. In contrast, ions that are produced at phases close to a zero-crossing of the electric field, where the returning electron has enough energy to ionize its parent ion, acquire large drift momenta, resulting in a final momentum distribution with a distinct minimum at zero [7].

The non-sequential double ionization was reported for all rare gas atoms [2,5,8]. However, a well-pronounced double peak structure in the longitudinal momentum distribution of the doubly charged ions was observed only for Ne, whereas for He and Ar a considerable fraction of ions was found to have small momenta [9]. This difference has been explained to be due to the contribution of another mechanism of electron-electron correlation, namely *r*ecollision induced *e*xcitation followed by *s*ubsequent field *i*onization (RESI) [9-11]. Here the returning electron excites the parent ion which is then ionized via tunneling in one of the subsequent cycles of the oscillating laser field (see Fig. 1). The latter step most likely occurs near one of the field maxima, resulting in small ion drift momenta and, thus, in a 'filling of the valley' in the momentum distributions. Estimations of the cross sections for collision-induced ionization/excitation [9] as well as the analysis of correlated electron momenta [10-12] showed that RESI plays a decisive role for He and Ar, whereas for Ne recollision induced electron-impact ionization clearly dominates.

The aforementioned enhancement of ion yields was observed also for cases where more than two electrons are ionized [2,8,13-15]. The description of multiple ionization represents even a greater challenge to theory, not only because of the increased number of reaction partners, but also due to the larger variety of the possible pathways leading to the same final state, including different combinations of sequential and non-sequential processes [16,17], and the possibility of multiple excitations. For triple ionization of Ar it was shown that the neglect of specific nonsequential channels leads to large discrepancies between measured and calculated yields of multiply charged ions [16]. However, yield measurements alone do not allow one to determine particular mechanisms explicitly. In this paper we present differential experimental data on multiple (up to fourfold) ionization of Ne and Ar, which allowed us to disentangle different multiple ionization mechanisms, since they populate different regions of the final state momentum space. We confirm our findings presenting the very first experimental results on double and multiple ionization with few – cycle (7 fs) laser pulses.

In order to identify the mechanisms contributing to multiple ionization, we measured the longitudinal momentum distributions of the recoil ions. To relate the observed ion momenta to certain multiple ionization pathways it is convenient to consider the individual emitted electrons since their sum momentum balances the ion momentum. For any combination of sequential and non-sequential processes, two classes of electrons can be distinguished: those emitted via field (tunnel) ionization ($e_T$) from the ground or excited state, and electrons emerging from a recollision process ($e_R$). The final momentum of an ion with charge state $n$ is given by:

$$p_{ion}^{(n)} = -\sum_{i=1}^{l} p_i(e_T) - \sum_{j=1}^{m} p_j(e_R), \quad n = l + m \quad . \tag{1}$$

As can be seen from Fig.1, electrons emitted via tunneling acquire a small drift momentum and, thus, give only a small contribution to the final sum momentum (1). Hence, the width of the ion momentum distribution is mainly defined by the contribution of electrons emitted simultaneously via a recollision process, which happens preferentially close to a zero-crossing of the oscillating electric field. Electrons which are set free exactly at a zero-crossing gain a maximum drift momentum of $2\sqrt{U_p}$ (where $U_p$ is the so-called ponderomotive potential, $U_p = E^2/4\omega^2$, E is the electric field strength and $\omega$ is the light frequency). Thus, an upper boundary for the most probable final longitudinal momentum of an n-charged ion can be estimated to be $p_{max} = 2n\sqrt{U_p}$, which corresponds to the case where all electrons are set free simultaneously during a recollision event. Any other sequence of processes like subsequent tunneling of two electrons followed by recollision-induced ionization as a last step will lead to considerably smaller momenta [17].

The measurements were performed using a new "reaction microscope" [18] designed to meet the specific requirement of the experiments with high-intensity lasers (see [11] for details). We used linearly polarized radiation of a Kerr-lens mode locked Ti:sapphire laser at 795 nm wavelength with 25 fs pulse width (FWHM). To generate few-cycle pulses they were spectrally broadened in a gas-filled hollow fiber and then compressed to 6-7 fs (FWHM) by chirped mirrors and a prism compressor. The laser beam was focused to a spot size of ~ 7 µm on the collimated supersonic gas jet in the ultra-high vacuum chamber ($2\cdot10^{-11}$ mbar). Created reaction fragments were guided to two position-sensitive channel plate detectors by weak electric (1V/cm) and magnetic (5G) fields applied along the laser polarization axis. From the time-of-flight and position on the detectors the full momentum vectors of the recoil ions and electrons were calculated.

In Fig. 2 the longitudinal momentum distributions of $Ne^{2-4+}$ ions obtained with 25 fs 800 nm laser pulses are presented. At an intensity of $1.5\cdot10^{15}$ W/cm$^2$ the distributions of both $Ne^{2+}$ and $Ne^{3+}$ exhibit the characteristic double-peak structure (Fig. 2a), a clear signature of recollision-induced non-sequential ionization. At higher intensity, $Ne^{2+}$ ions are produced sequentially via two independent tunnel ionization events, and the corresponding momentum distribution is a Gaussian with a maximum at zero (Fig. 2b). The momentum distributions of $Ne^{3+}$ and $Ne^{4+}$ ions at the same intensity, however, still exhibit a clear double-peak structure, with almost no ions created at zero momentum (Fig. 2b, 2c). The spectra spreads even slightly beyond the classical cut-offs $p_{max}$ represented by the arrows. Such high recoil ion momenta can be obtained only when all of the emitted electrons participate in the recollision event. Contributions beyond $p_{max}$ reflect a probability for the returning electron to be backscattered on the parent ion and, thus, getting a momentum higher then the maximum drift momentum. However, the probability of the backscattering is rather small, and a region of lower momenta is kinematically favoured [17].

Thus, non-sequential enhancement of double, triple and fourfold ionization of Ne proceeds mainly via a direct recollision-induced process involving up to four electrons. Very surprisingly, the dominant pathway of the $Ne^{3+}$ and $Ne^{4+}$ ions production does not change when the double ionization passes from non-sequential to sequential regime. This highly correlated electron emission occurs on a sub-laser cycle time scale, and all processes that occur on longer time scales seem to give a negligible contribution.

Inspecting the momentum distributions of multiply charged Ar ions created by 25 fs laser pulses (Fig. 3), we observe a completely different behaviour. In the intensity regime where the yield measurements showed a qualitatively similar 'non-sequential' enhancement for Ne and Ar [8,16], striking differences in the momentum distributions are observed. The $Ar^{3+}$ distributions exhibit only a shallow minimum at zero, whereas for $Ar^{4+}$ ions we do not observe a double-peak structure at all, pointing to other mechanisms to be of importance. Considering the width of the spectra, one can see that for Ar the data lie well within the $2n\sqrt{U_p}$ limits. Moreover, at the intensity where both double and triple ionization of Ar occurs predominantly non-sequentially (Fig. 3a), the momentum distributions of doubly and triply charged ions are very close: the $Ar^{3+}$ spectrum is only slightly broader (compare with Fig. 2a). The most likely mechanism of triple ionization thus involves the sequential production of $Ar^{2+}$ with subsequent recollision. Then, the contribution of the first tunnelled electron in the momentum sum (1) will lead only to the slight broadening of the final ion momentum distribution.

Since the ion-momentum distributions show very similar behaviour for double and triple ionization, one might assume that similar mechanisms are active, and differences in the shape of multiple ionization spectra for Ne and Ar might be attributed to the different importance of the RESI mechanism, as it has been convincingly demonstrated for double ionization [9]. To confirm this, we present the first experimental data on multiple ionization by few-cycle laser pulses. The RESI pathway should be suppressed when few-cycle pulses are used. Indeed, the probability of the ion excited by the rescattered electron to be ionized depends on the intensity of the laser field at the subsequent maxima. In contrast, any direct (e,ne) process is not influenced at all by the pulse length if it occurs at the first return of the electron. Comparing the results obtained with 25 and 7 fs pulses (Fig. 4) we observe that the momentum distribution for double ionization of Ne, which is dominated by recollision-induced (e,2e) process, do not depend on the pulse length, whereas for all three charge states of Ar there is a significant modification of the spectra. For 7 fs pulse RESI contributions resulting in ions with small longitudinal momenta are suppressed, and the characteristic double-hump structure is restored, making the shape of the spectra closer to those of Ne. A similar depletion of the region around zero momentum could be also caused by the suppression of the sequential ionization from the ground state. However, the contribution of purely sequential production of $Ar^{2+}$, $Ar^{3+}$ and $Ar^{4+}$ at the intensities of Fig. 4b, 4c, and 4d, respectively, is still a small fraction of the total yield [16], and thus, can not be responsible for the changes observed.

In conclusion, we found that essentially only one pathway based on the "instantaneous" interaction between up to four electrons dominates multiple electron removal from Ne. Surprisingly, triply and fourfold charged ions are predominately produced by highly correlated sub-cycle few-electron emission even at those intensities where the dominating mechanism of double ionization does not involve electron correlation. In Ar, instead, different types of less correlated multi-electron pathways involving the RESI mechanism and cascades of sequential and non-sequential processes are important. Especially interesting is the fact that the atomic structure plays an important role even at intensities well above $10^{15}$ W/cm$^2$ by efficiently "selecting" the most effective dynamical pathway to multiple ionization, i.e. the coupling between several atomic electrons and the intense laser field.

The demonstrated possibility to influence many-electron dynamics by changing the laser pulse parameters can allow for a control over various field-induced phenomena. In particular, multiple ionization of Ne, which occurs essentially via only one clear pathway, represents a sensitive tool to study the carrier envelope phase effects [19], since the returning electron can have enough energy to induce multiple ionization only if it was born at certain initial phase. Recently the production of ~ 5 fs laser pulses with stabilized carrier envelope phase was reported [20], and use of such pulses can allow one to manipulate the details of the recollision process.

**Figure 1**. Mechanisms of strong-field ionization. If the electric field changes with time as $E(t) = E_0 \sin \omega t$, an electron which is set free at the time t acquires a final drift momentum of $P_{drift}(t) = q/\omega\, E_0 \cos(\omega t)$. Thus, an electron created near the maximum of the field *(1)* emerges with zero momentum after the laser pulse has gone. In sequential double ionization *(4)* tunneling happens twice leading to small momenta of the produced ion, which balances the electron sum-momentum. If an electron tunnels out shortly after the field maximum it can recollide with its parent ion where it may either excite *(2)* or directly knock out *(3)* another atomic electron. In the latter case both electrons acquire a large net momentum of up to $2\sqrt{U_P}$ each. The corresponding final ion momentum displays clear maxima. In cases where an electronic excitation occurs during recollision only the slowed down projectile electron acquires a namable momentum from the laser field, while the excited electron, which is field ionized at a later time, appears as a slow electron. This pathway (RESI) gives rise to ion-momentum distributions slightly broader than for the sequential case.

**Figure 2.** Longitudinal momentum distributions of neon ions obtained with 25 fs 800 nm laser pulses. The laser peak intensity in PW/cm$^2$ as well as the ion charge states are indicated in the each figure. Arrows indicate momenta of $\pm 2n\sqrt{U_P}$ with n = 3 for the upper two panels and n = 4 for the lowest.

**Figure 3.** Same as figure 2 for argon. Arrows indicate momenta of $\pm 2n\sqrt{U_P}$ with n = 3 for the upper two panels and n = 4 for the lowest.

**Figure 4.** Longitudinal momentum distributions of recoil ions obtained with 25 fs (squares) and 7 fs (solid line) 800 nm laser pulses. (a) Ne$^{2+}$, 0.8 PW/cm$^2$. (b) Ar$^{2+}$, 0.5 PW/cm$^2$. (c) Ar$^{3+}$, 0.8 PW/cm$^2$ (d) 1.0 PW/cm$^2$, Ar$^{4+}$.

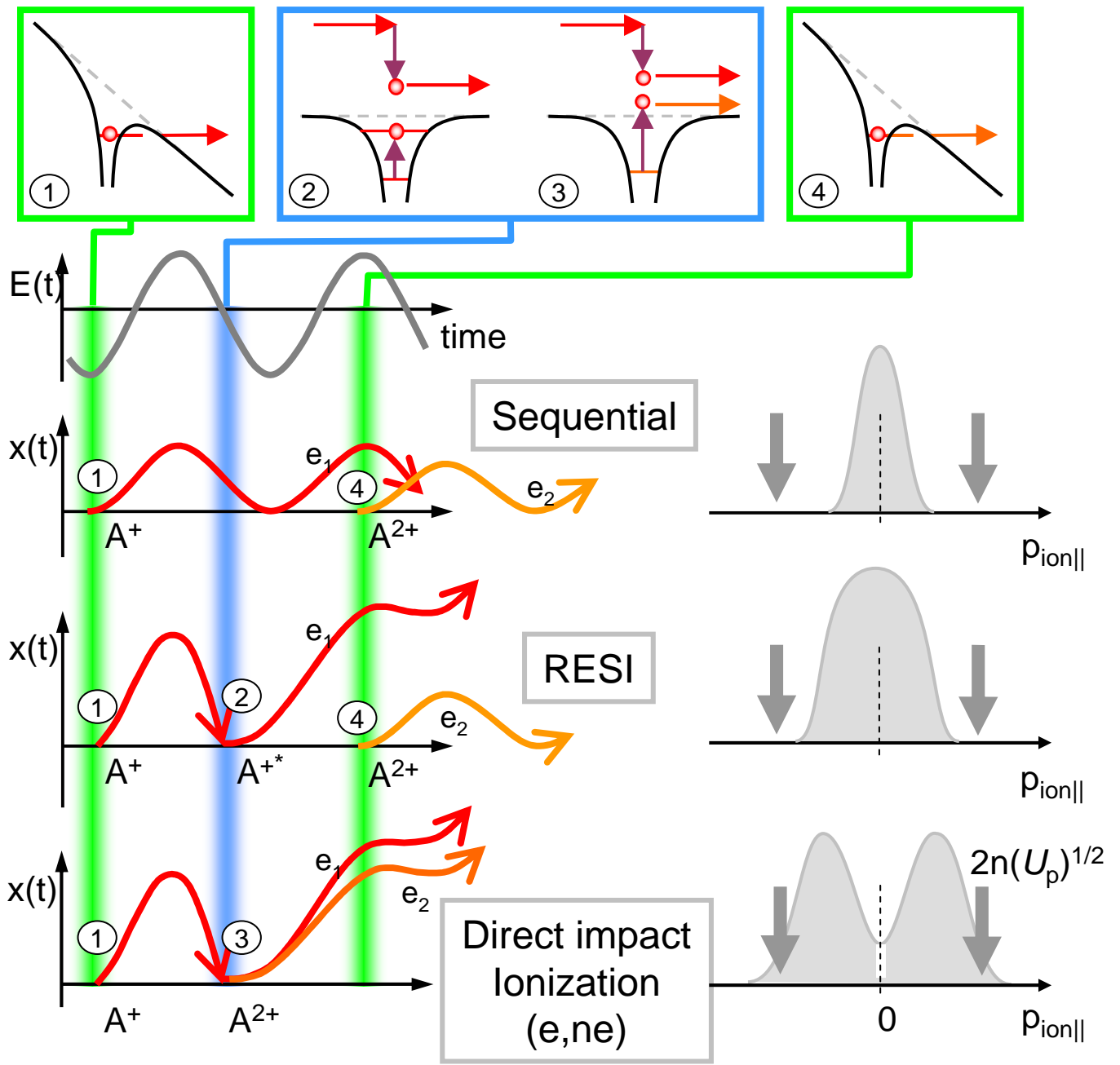

Figure 1 (color online)

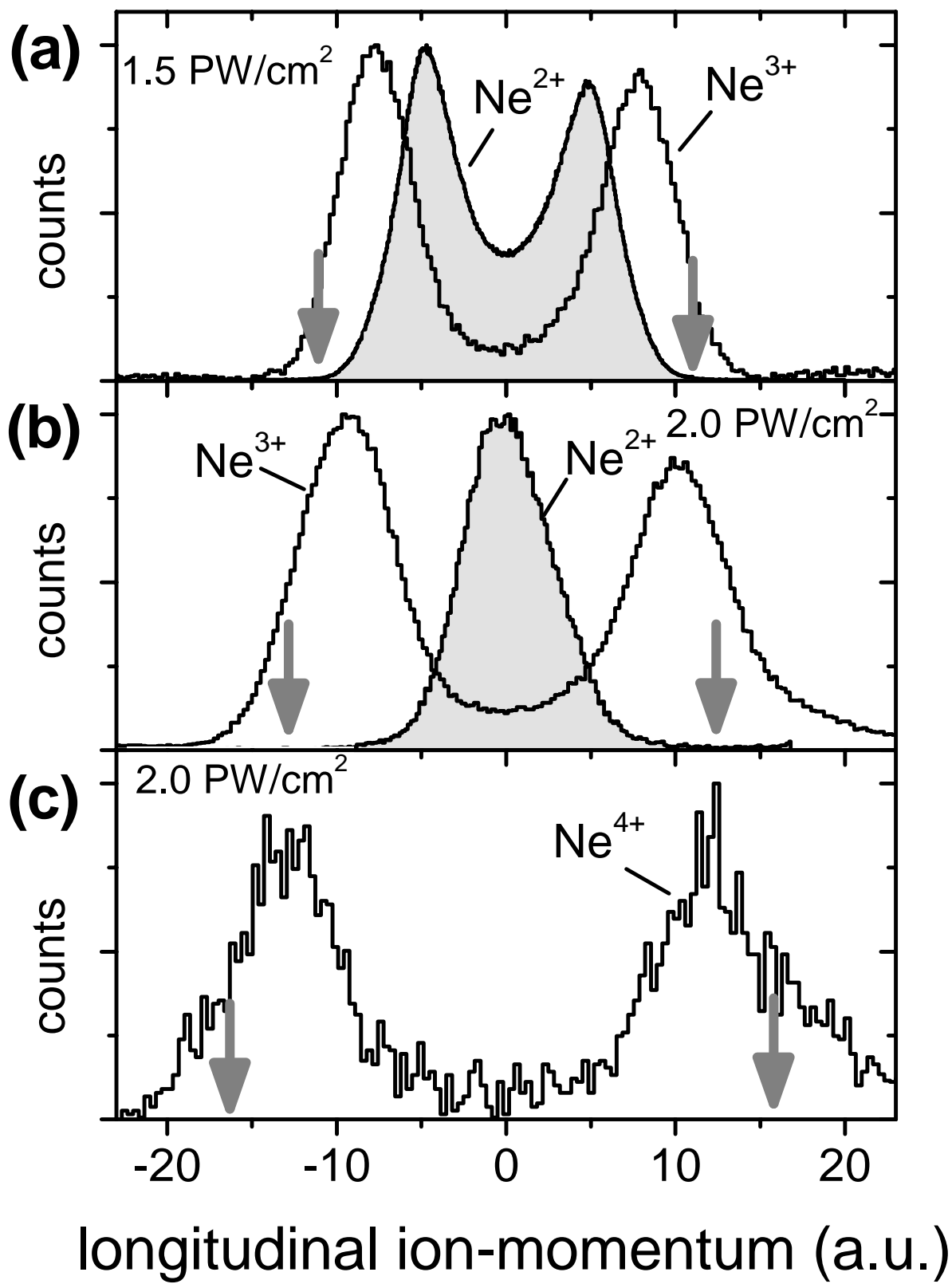

Figure 2

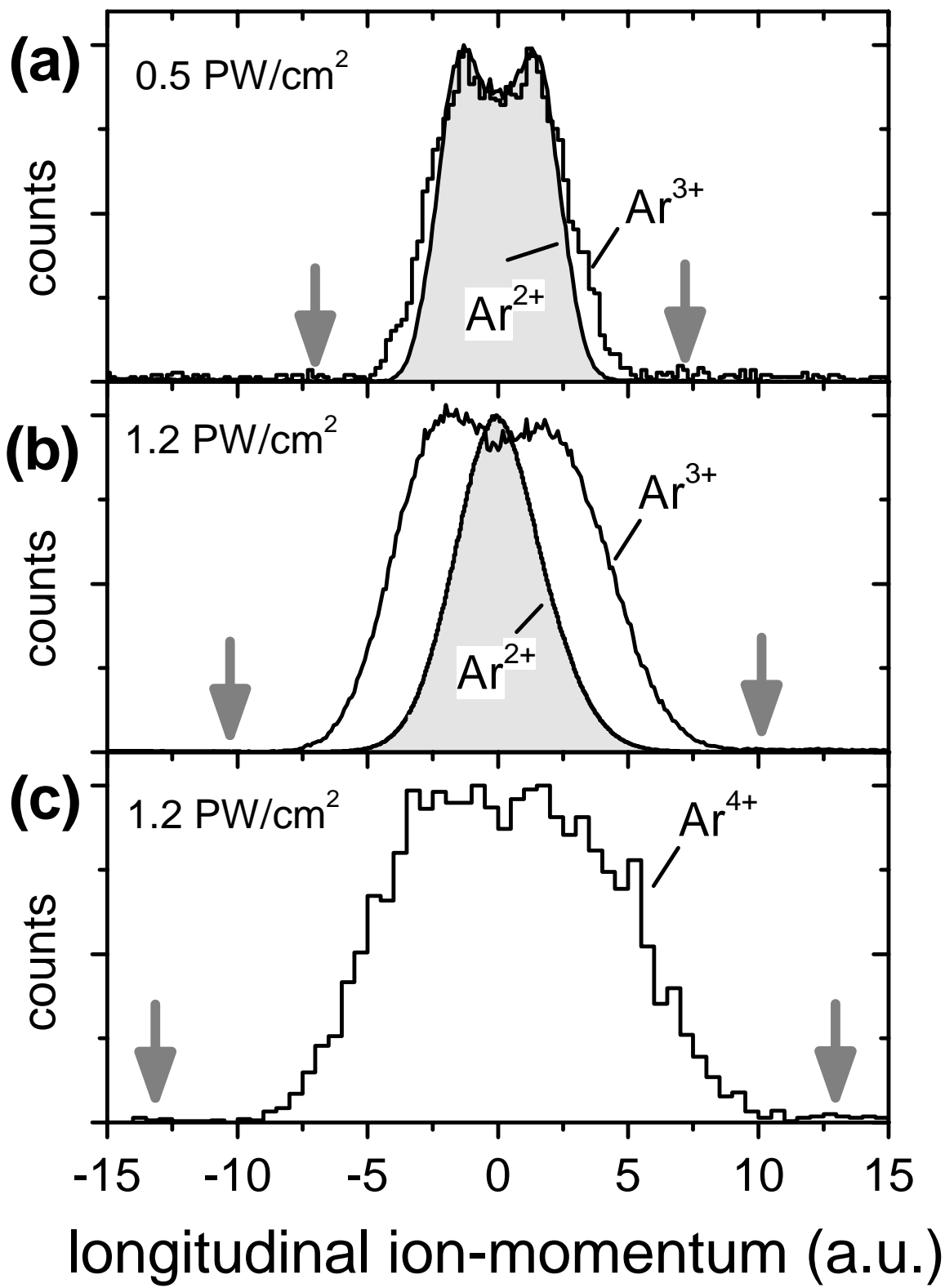

Figure 3

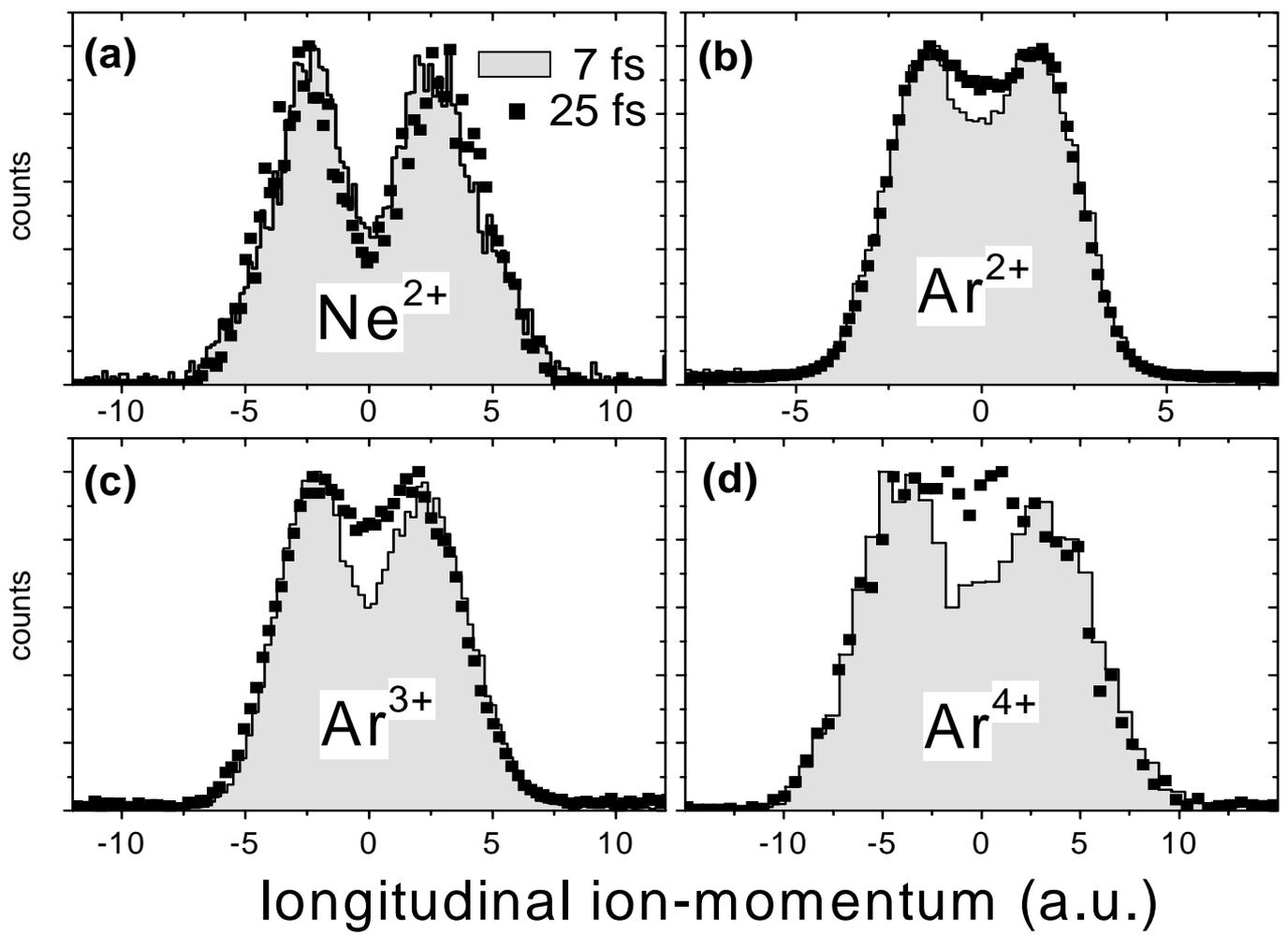

Figure 4